\documentclass[onecolumn,showpacs,floats,floatfix,aps,prl,groupedaddress]{revtex4}
\usepackage{graphicx}
\usepackage{url}
\usepackage{amsmath}
\usepackage{subfigure}

\begin{document}

\title{Dynamical Analysis of the Structure of Neutron Star Critical Collapses}

\author{M.-B. Wan$^{2}$, K.-J. Jin$^{2}$, W.-M. Suen$^{1,2}$}

\affiliation{
{$^1$}Department of Physics, The University of 
Hong Kong, Hong Kong\\
{$^2$}McDonnell Center for the Space Sciences, 
Department of Physics, Washington University, St. Louis, 
Missouri 63130}

\date{\today}

\begin{abstract}
Jin \textit{et~al} reported that axisymmetric simulations of NS-like objects
with polytropic EOS undergo critical gravitational collapse. As the critical 
collapse observed via fine-tuning of the adiabatic index $\Gamma$, they 
conjecture that critical phenomena may occur in realistic astrophysical 
scenarios. To clarify the implications this numerical observation has on 
realistic astrophysical scenarios, here, we perform dynamical analysis on the 
structure of the critical collapse observed in the former work. We report the 
time scales and oscillation frequencies exhibited by the critical solution and 
compare these results with values obtained from analytic perturbative mode 
analysis of equilibrium TOV configurations. We also establish the universality of the critical 
solution with respect to a 1-parameter family of initial data as well as the phase space
manifold of the critical collapse.


\end{abstract}
\maketitle

\paragraph*{\bf Sec.1. Introduction.}

Critical phenomena in gravitational collapse was discovered by Choptuik 
almost a decade ago \cite{Choptuik}. Since then, there has been much 
development in establishing the theory in its mathematical aspects. Past 
studies have considered the assumption of both spherically 
symmetric and axisymmetric systems, and several matter models eg neutron 
stars, perfect fluids, real scalar fields, 2D Sigma models, SU(2) 
Yang-Mills and Skyrme models, were considered in these studies. However, 
thus far, none have considered axisymmetric systems of neutron stars. The 
neutron star models that have been considered \cite{Noble} employ 
spherical symmetry and Type II critical phenomena were found in these 
studies. 

Jin \textit{et~al} \cite{Jin} carried out axisymmetric simulations of 
neutron stars and found Type I critical phenomena ie. black holes with a 
mass gap form in the black hole phase. The axisymmetry in these studies enable 
high-resolution finite differencing that contributes to the 
observation of the fine structure of critical phenomena when n-parameters 
of the system are fine-tuned. In this study, Jin \textit{et~al} tuned the 
central densities, the boost velocities, the initial separations and the 
polytropic index of the EOS of the NSs one at a time. Varying solely the central densities of the NSs corresponds to varying the 
baryonic masses of the NSs. Configurations with supercritical baryonic masses result in black holes while those with subcritical 
baryonic masses result in NSs. The system is carried away from the critical 
solution by the 1 unstable mode. Even in the absence of a scale in the Einstein field equations, in Type I critical 
phenomena, the time of departure from the critical solution scales with respect to constant 
powers of the distance of the initial data from the critical threshold.

As critical phenomena is found in the above study when the polytropic 
index of the EOS is tuned, it is conjectured that NS-like objects 
undergoing a slow cooling process, accretion and angular momentum loss may 
exhibit critical phenomena. In order to clarify the astrophysical 
implications this may entail, we extract the relevant time scales from the critical solution. Since exact critical 
solutions in principle stay on the critical point as time proceeds to infinity, we perform a perturbative mode analysis and  
compare the mode frequencies of the critical solution with frequencies of specifically non-radiating modes for TOV configurations 
with the same baryonic mass. 

We next determine the universality of these solutions with respect to a 1-parameter family of 
initial data, by constructing packets of matter using the polytropic EOS 
($p = K \rho^{\Gamma}$) with $K=80$ and $\Gamma=2.0$, whose distributions 
are characterized by a Gaussian rotated about the axis of axisymmetry. We 
perform collision simulations similar to that done in \cite{Jin} using 
these packets of matter and observe that the critical solution found there 
constitutes a universal attractor. We suggest strong evidence that the critical point found in Jin \textit{et~al} is a limit
cycle rather than a fixed point and present phase diagrams using
appropriate evolution variables in the 3+1 split. All parameters and variables presented in this work are in $G=c=1$ units.

\paragraph*{\bf Sec.2. Perturbative Mode Analysis.} 

\begin{figure}
\includegraphics[scale=0.85]{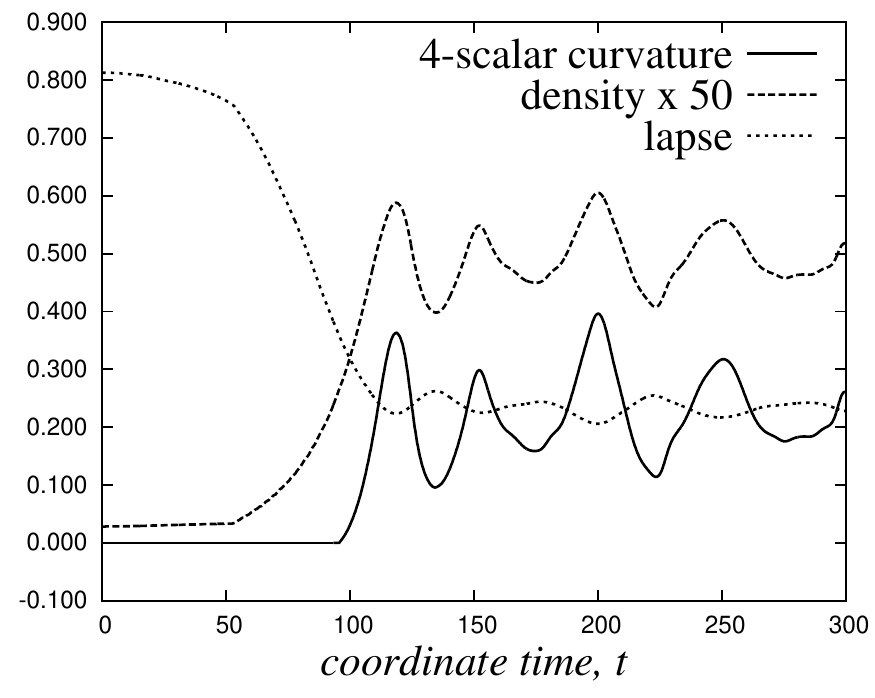}
\caption{}
\end{figure}
In this analysis, we first obtain the time scale of attraction of the system toward the critical solution and the dominant frequencies 
of the oscillations during the period of the critical solution. Fig. 1 shows the evolution of the lapse, density and 4-scalar curvature at 
the center of collision. We first obtain, via a Fourier analysis, the dominant frequencies of the oscillations of these 3 evolution 
variables. Fig. 2 indicates 2 dominant frequencies, namely, $\omega_{1}=0.15M_{\odot}$ and $\omega_{2}=0.23M_{\odot}$, which correspond 
to oscillation periods $T_{1}=0.21ms$ and $T_{2}=0.13ms$ respectively, that are universal across all 3 variables. Using the dominant 
frequencies, we then fit the entire critical solution with the following equations:\\
lapse:
\begin{equation}
\alpha=\alpha_{\infty}+ae^{-\lambda(t-90)}+b(\cos(\omega_{1}(t-90)+c)+ratio_{\alpha}\cos(\omega_{2}(t-90)+d))                        
~\end{equation}
density:
\begin{equation}
\rho=(\frac{1}{\frac{1}{\rho_{\infty}}+ae^{-\lambda(t-60)}})-b(\cos(\omega_{1}(t-60)+c)+ratio_{\rho}\cos(\omega_{2}(t-60)+d))
~\end{equation}
4-scalar curvature:
\begin{equation}
R=(\frac{1}{\frac{1}{R_{\infty}}+ae^{-\lambda(t-100)}})-b(\cos(\omega_{1}(t-100)+c)+ratio_{R}\cos(\omega_{2}(t-100)+d))
~\end{equation}
where a, b, c and d are fit constants, $\alpha_{\infty}$, $\rho_{\infty}$ and $R_{\infty}$ are respectively the averages of the 
lapse, density and 4-scalar curvature throughout the oscillation phase of the critical solution, $ratio_{\alpha}=0.2$, 
$ratio_{\rho}=ratio_R=0.4$ are the ratio of the amplitudes of $\omega_2$ over $\omega_1$ obtained from the Fourier transforms of the 
respective evolution variables. The fit results are shown in Fig. 3. From the fit, we find that $\lambda=0.09$, which corresponds to the 
attraction time, $t=0.05ms$, of the system toward the critical solution, is universal across the 3 evolution variables. We further observe 
that the attraction time is similar to the departure time found in \cite{Jin}. This indicates the existence of a complex pair of eigenvalues 
of the growing unstable mode when linearization in a small neighbourhood of the critical solution is performed. We note that these time 
scales are of an order of magnitude smaller than the cooling time scales of NSs commonly reported in the astronomy and astrophysics 
literature. Therefore, there is a high possibility that real astrophysical systems of cooling NSs pass through the threshold of 
critical collapse. Further, systems that undergo slow accretion and angular momentum loss with time scales larger than that reported 
here for the NS critical solution may also pass through the threshold and exhibit critical behavior.
\begin{figure} 
\includegraphics[scale=0.8]{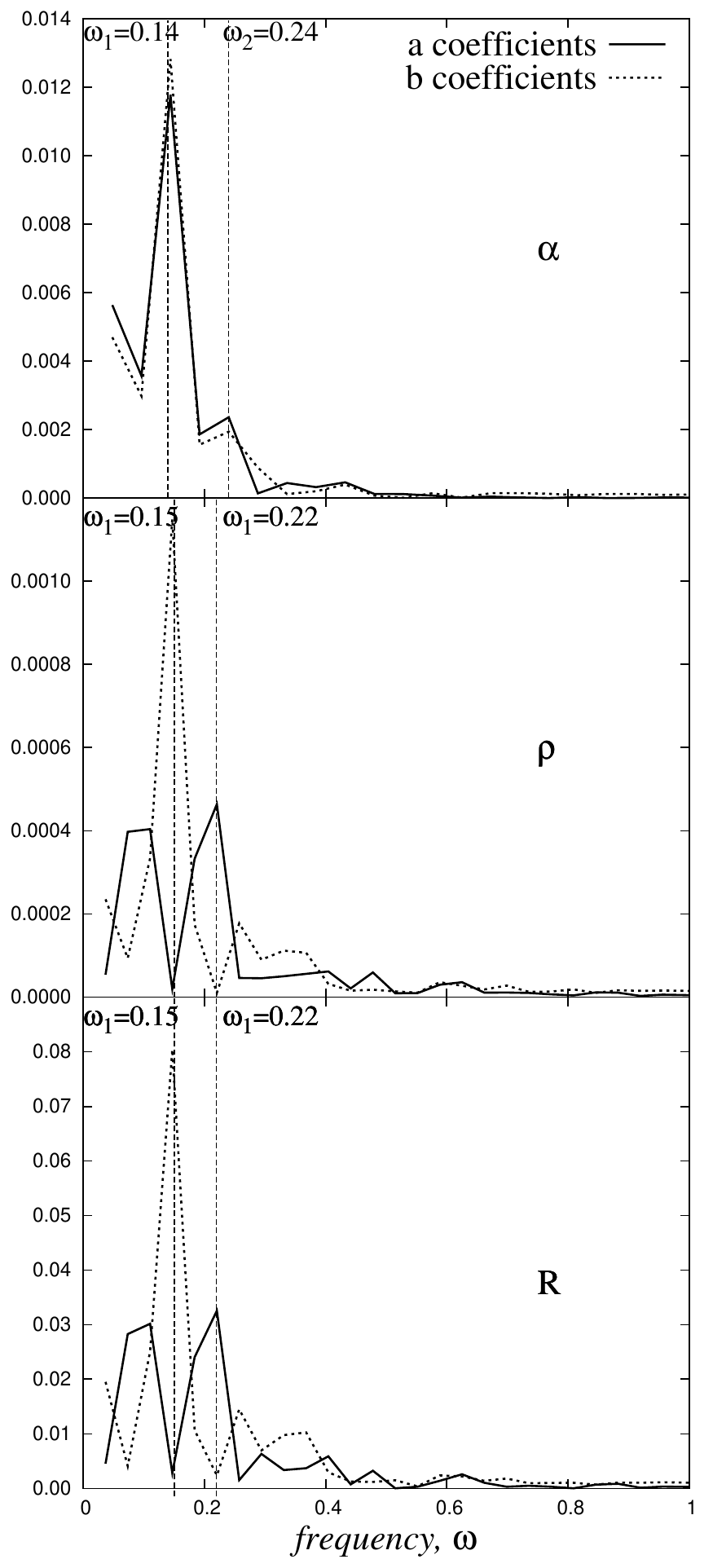}
\caption{}
\end{figure}
\begin{figure}
  \begin{center}
    \begin{tabular}{c}
      \scalebox{0.8}{\includegraphics{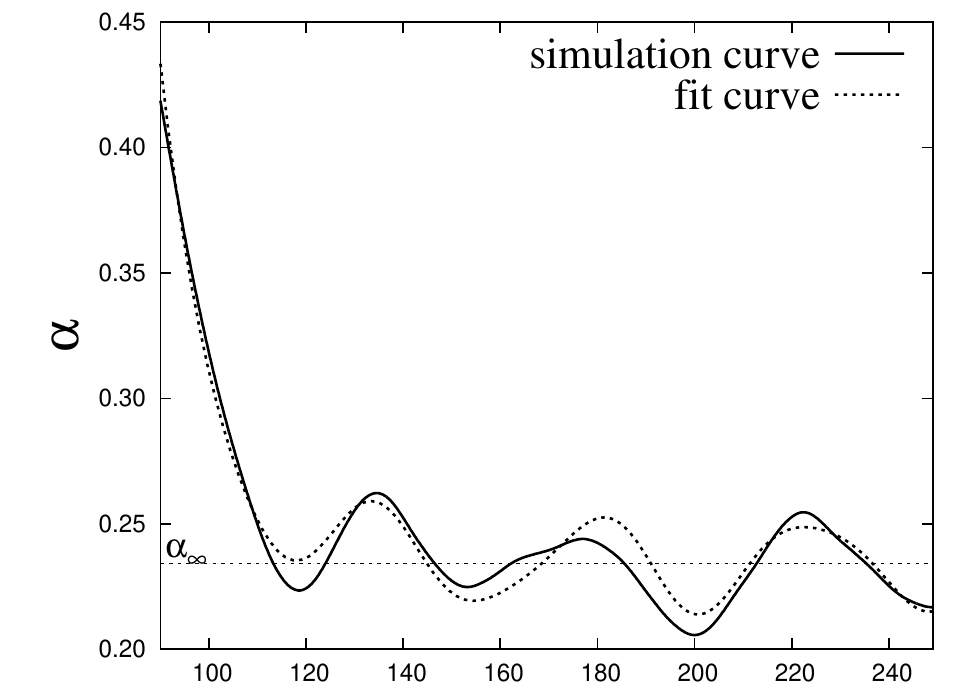}} \\
      \scalebox{0.8}{\includegraphics{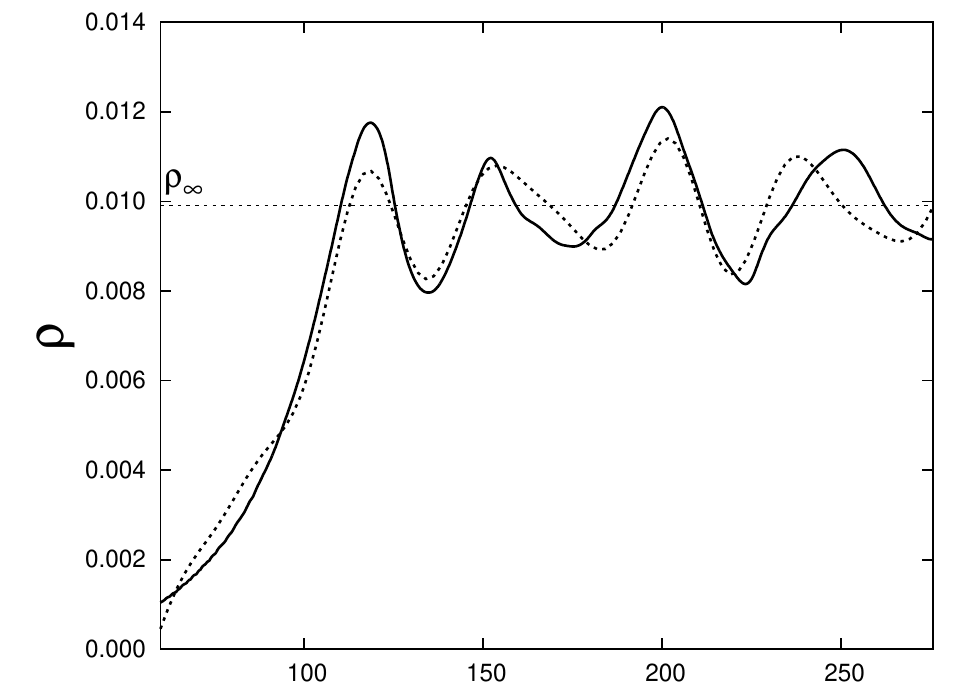}} \\
      \scalebox{0.8}{\includegraphics{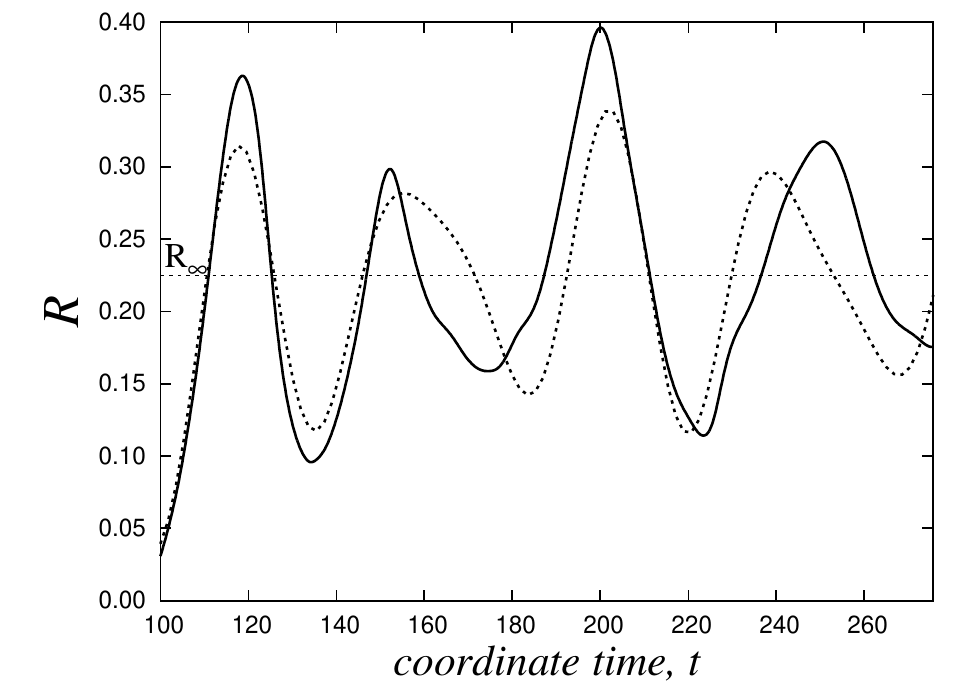}}
    \end{tabular}
    \caption{}
    \label{3}
  \end{center}
\end{figure}

We next perform a perturbative mode analysis on TOV configurations that have the same baryonic mass as the object that undergoes 
critical collapse, namely at baryonic mass $M_b=1.5393M_{\odot}$. We note that the exact critical solution is non-radiative in 
principle, and therefore, we focus on the l=0,1 modes. Using the following perturbation metric on a TOV background \cite{Campolatarro}:
\begin{equation}
\begin{pmatrix}
H_{0}e^{\nu}Y^{1}_{m}  & i\omega H_{1}Y^{1}_{m} & h_{0}\Psi^{1}_{m\theta} & h_{0}\Psi^{1}_{m\phi}  \\
i\omega H_{1}Y^{1}_{m}  & H_{2}e^{\lambda}Y^{1}_{m} & h_{1}\Psi^{1}_{m\theta} & h_{1}\Psi^{1}_{m\phi} \\
0 & 0 & r^2(G-K)\Psi^{1}_{m\theta\theta} & r^2(G-K)\Psi^{1}_{m\theta\phi} \\
0 & 0 & r^2(G-K)\Psi^{1}_{m\phi\theta} & r^2(G-K)\Psi^{1}_{m\phi\phi}
\end{pmatrix}
~\end{equation}
with a gauge choice that leaves the exterior solution invariantly spherically symmetric, namely $G=K\equiv 0$, and $h_{0}=h_{1}\equiv 
0$, we obtain the following perturbed line elements:\\
l=0 mode:
\begin{equation}
ds^2=e^{\nu}(H_{0}-1)dt^2+e^{\lambda}(H_{2}+1)dr^2+r^2(d\theta^2+\sin^2\theta d\phi^2)
~\end{equation}
l=1 mode:
\begin{equation}
ds^2=e^{\nu}(H_{0}Y^{1}_{m}-1)dt^2+2i\omega H_{1}Y^{1}_{m}dtdr+e^{\lambda}(H_{2}Y^{1}_{m}+1)dr^2+r^2(d\theta^2+\sin^2\theta d\phi^2).
~\end{equation}
We further refer to the various past established literature \cite{Lindblom}, \cite{Misner}, \cite{Kokkotas} and 
\cite{Bardeen}, reconcile the differing conventions and confirm their results with our mode analysis. 

For the l=0 mode, we denote the fluid perturbation as:
\begin{equation}
\xi=-r^{-2}e^{-\lambda/2}W.
~\end{equation}
Assuming harmonic dependence on time for the fluid perturbation as the only degree of freedom in this calculation, we thus write:
\begin{equation}
W(r,t)=W(r)e^{i\omega t}.
~\end{equation}
We then solve the following equations:\\
$\delta G_{1}^{0}=8\pi T_{1}^{0}$:
\begin{equation}
r^{-1}e^{-\lambda}H_{2,t}-8\pi(\rho+p)r^{-2}e{-\lambda/2}W_{,t}=0
~\end{equation}
$\delta G_{1}^{1}=8\pi T_{1}^{1}$:
\begin{equation}
2r^{-1}e^{-(\lambda+\nu}H_{1,t}-r^{-1}e^{-\lambda}H'_0-r^{-2}e^{-\lambda}(1+r\nu ')H_2-8\pi\gamma p(r^{-2}e^{-\lambda/2}W'-\frac{H_2}{2})
-8\pi r^{-2}e^{-\lambda/2}p'W=0
~\end{equation}
projection of $\delta(T^{\mu}_{1;\mu})=0$ orthogonal to $\b{u}$:
\begin{equation}
(\rho+p)\ddot{\xi}e^{2(\lambda-\nu}-\frac{(\rho+p)}{2}H'_0+(\delta\rho+\delta p)\nu '+\delta p'=0
~\end{equation}
and obtain the following 1st-order ODE:
\begin{equation}
\begin{split}
(\frac{p'}{\rho '})'(r^{-2}e^{-\lambda/2}W)''+((\frac{p'}{\rho '})'-Z
+4\pi r\gamma pe^{\lambda}-\frac{\nu '}{2})(r^{-2}e^{-\lambda/2}W)'
+(\frac{\nu '}{2}+\frac{2m}{r^3}e^{\lambda}-Z'-4\pi(\rho+p)Zre^{\lambda}
+\omega^2e^{\lambda-\nu})\\
(r^{-2}e^{-\lambda/2}W)=0
\end{split}
~\end{equation}
where $Z=(\frac{p'}{\rho '})(\frac{\nu'}{2}-2r^{-1})$, imposing the boundary condition that the pressure perturbation goes to zero at 
the TOV star surface, which, according to \cite{Bardeen}, is equivalent to:
\begin{equation}
(r^{-2}e^{-\lambda/2}W)'=(\gamma R)^{-1}(4+(\frac{M}{R})e^{\lambda}W+\omega^2(\frac{R^3}{M})e^{-\nu})(r^{-2}e^{-\lambda/2}W),
~\end{equation}
where M is the total ADM mass of the TOV star. From a simple shoot-and-match calculation of a finite-differenced TOV configuration as 
mentioned above, we find that the fundamental frequency is $\omega_{1p}=6.858\times 10^{-3}M_{\odot}$.

Similarly, for the l=1 mode, we have:
\begin{equation}
\left.\begin{matrix}
\xi^{r}=-r^{-2}e^{-\lambda/2}WY^{1}_{m}\\ 
\xi^{\theta}=\frac{V}{r^2}\Psi^{1\theta}_{m}\\
\xi^{\phi}=\frac{V}{r^2}\Psi^{1\phi}_{m} 
\end{matrix}\right\},
~\end{equation}
as the fluid perturbations, and write their corresponding harmonic dependences:
\begin{equation}
\left.\begin{matrix}
H_{0}(r,t)=H_{0}(r)e^{i\omega t}\\ 
H_{1}(r,t)=H_{1}(r)e^{i\omega t}\\
H_{2}(r,t)=H_{2}(r)e^{i\omega t}\\
W(r,t)=W(r)e^{i\omega t}\\
\end{matrix}\right\}.
~\end{equation}
We solve the following equations:\\
$\delta G_{0}^{0}=8\pi T_{0}^{0}$:
\begin{equation}
H'_2+(r^{-1}(1+e^{\lambda})-\lambda'+4\pi re^{\lambda}(\rho+p))H_2=8\pi 
re^{\lambda}[(\rho+p)r^{-2}e^{-\lambda/2}W'+r^{-2}e^{-\lambda/2}\rho'W+2r^{-2}(\rho+p)V
~\end{equation}
$\delta G_{1}^{j}=8\pi T_{1}^{j}$:
\begin{equation}
rH'_0=-(\frac{r\nu'}{2}-1)H_0-(\frac{r\nu'}{2}+1)H_2+iwre^{-\nu}H_1
~\end{equation}
$\delta(T^{\mu}_{j;\mu})=0$:
\begin{equation}
-(\rho+p)e^{-\nu}\omega^2 V=-p'r^{-2}e{-\lambda/2}W-p\gamma r^{-2}e^{-\lambda/2}W'+\frac{\gamma p}{2}H_2+\frac{(\rho+p)}{2}H_0-2\gamma 
pr^{-2}V,
~\end{equation}
to obtain a 3rd-order ODE:
\begin{equation}
\left.\begin{matrix}
H_2=-r^{-1}H_1+r^{-1}8\pi(\rho+p)e^{\lambda/2}W\\
16\pi r^2\omega^2(\rho+p)e^{\lambda-\nu}V=-8\pi(\rho+p)\nu'e^{\nu/2}W+(\nu'-2r\omega^2 e^{-\nu})H_1+(16\pi r^2\rho 
e^{\lambda}-3r\lambda')H_0\\
H'_1=r^{-1}(\frac{r}{2}(\lambda'-\nu')-e^{\lambda})H_1+r^{-1}8\pi(\rho+p)e^{3\lambda/2}W-16\pi(\rho+p)e^{\lambda}V\\
rH'_0=-(\frac{r\nu'}{2}-1)H_0-(\frac{r\nu'}{2}+1)H_1-\omega^2 re^{-\nu}H_1\\
8\pi p\gamma e^{\lambda/2}W'=-r\omega^2 e^{-\nu}H_1-(\frac{r\nu'}{2}-4\pi p\gamma r^2 
e^{\lambda})H_2-(1-e^{\lambda}-\frac{r\nu'}{2})H_0-8\pi p'e^{\lambda/2}W-16\pi p\gamma e^{\lambda}V
\end{matrix}\right\}.
~\end{equation}
We then impose the boundary conditions that the space-time perturbation functions vanish at the TOV star surface whilst at the center of 
the star, the perturbation functions approach a finite value, namely:
\begin{equation}
\left.\begin{matrix}
H_0(R)=H_1(R)=0\\
\lim_{r\to 0}\frac{W}{r^2}=w\\
\lim_{r\to 0}\frac{H_0}{r^2}=h\\
\lim_{r\to 0}\frac{H_1}{r^2}=-8\pi(\rho_c+p_c)w\\
\end{matrix}\right\} 
~\end{equation}
where $w$ and $h$ are arbitrary constants. Via a variational principle calculation following \cite{Lindblom}, the fundamental frequency 
for this mode is found to be $\omega_{2p}=1.813\times 10^{-2}M_{\odot}$. 

We note that $\omega_{1p}$ and $\omega_{2p}$ are both of 1 and 2 orders of magnitude smaller than $\omega_1$ and $\omega_2$. Since 
$\omega_{1p}$ and $\omega_{2p}$ are the fundamental frequencies of the non-radiating modes of equilibrium configurations, our mode 
analysis confirms that the configuration that undergoes critical collapse is far from equilibrium, and thus a perturbative mode analysis 
has to be performed on non-equilibrium rather than on TOV background space-times in order to determine the nature of the oscillation 
frequencies exhibited by the critical solution.

\paragraph*{\bf Sec.3. Universality.}

We note that the initial data constructed by varying the boost velocity, initial separation and polytropic EOS are very close to each 
other and are within the perturbative region of the NS critical solution found in the former study. The universality of the critical 
solution with respect to a 1-parameter family of initial data is thus considered unclear \cite{Gundlach}. Therefore, in this section, we 
present a completely different family of initial data where the matter field consists of packets of matter whose densities are 
characterized by Gaussian distributions rotated about the axis of axisymmetry and a polytropic EOS ($p = K \rho^{\Gamma}$) with $K=80$ and 
$\Gamma=2.0$, where the tuning of the height and width of the Gaussian density distributions produce a 1-parameter family of initial data. 
Using the same numerical setup as in \cite{Jin}, namely $\Gamma$-freezing shift and 1+log slicing 3+1 BSSN evolution on a similar grid, 2 
Gaussian packets are boosted to a head-on collision and the behavior of the axisymmetric object formed is observed. Although the system 
is in a non-equilibrium state at the initial time, we have checked that their instability time scale, which is described by the 
oscillation time scale of a single Gaussian packet about an equilibrium TOV configuration, is much larger than the time 
scale of their merging and collapse, that the effect of the instability can be considered negligible. 
\begin{table}[ht]
\centering
\begin{tabular}{ccccc}
\hline\hline
Configuration & $M_b$ & $\rho_c$ & $d$ & $v_z$ \\ [0.5ex]
\hline
NS & 1.5 & 0.00056595 & 27.6 & 0.15 \\
1 & 1.5 & 0.00038698 & 29.6 & 0.1 \\
2 & 1.5 & 0.00039192 & 29.6 & 0.1 \\
3 & 1.6 & 0.00055501 & 27.6 & 0.12 \\
4 & 1.6 & 0.00064912 & 27.6 & 0.12 \\ [1ex]
\hline
\end{tabular}
\label{table:t1}
\caption{$M_b$ is the baryonic mass, $\rho_c$ is the NS central density/height of the Gaussian density distribution at $t=0$, $d$ is 
the center-to-center separation between the NSs/Gaussian packets, and $v_z$ is the boost velocity of the NSs/Gaussian packets along the 
z-direction of the grid.} 
\end{table}
\begin{figure}
\includegraphics[scale=0.85]{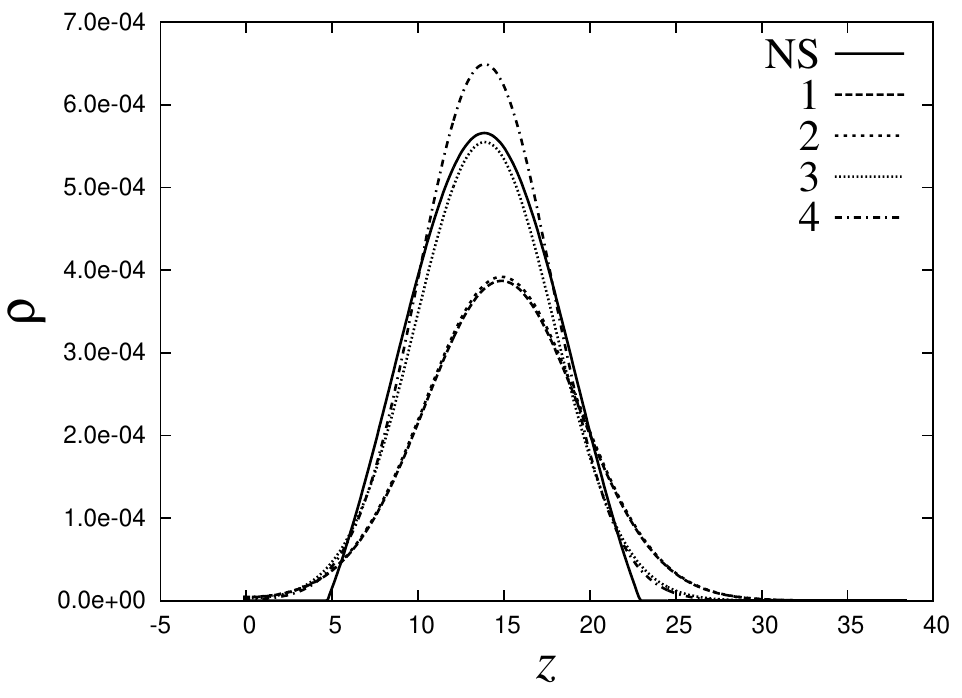}
\label{fig:4}
\caption{The Gaussian distributions are constructed so as to maintain the baryonic mass of the system as a constant. Therefore, when 
the Gaussian heights are increased, their widths are correspondingly decreased. Configurations 1 and 2 are less compact distributions 
whereas that of 3 and 4 are more compact.} 
\end{figure}   
\begin{figure}
\includegraphics[scale=0.8]{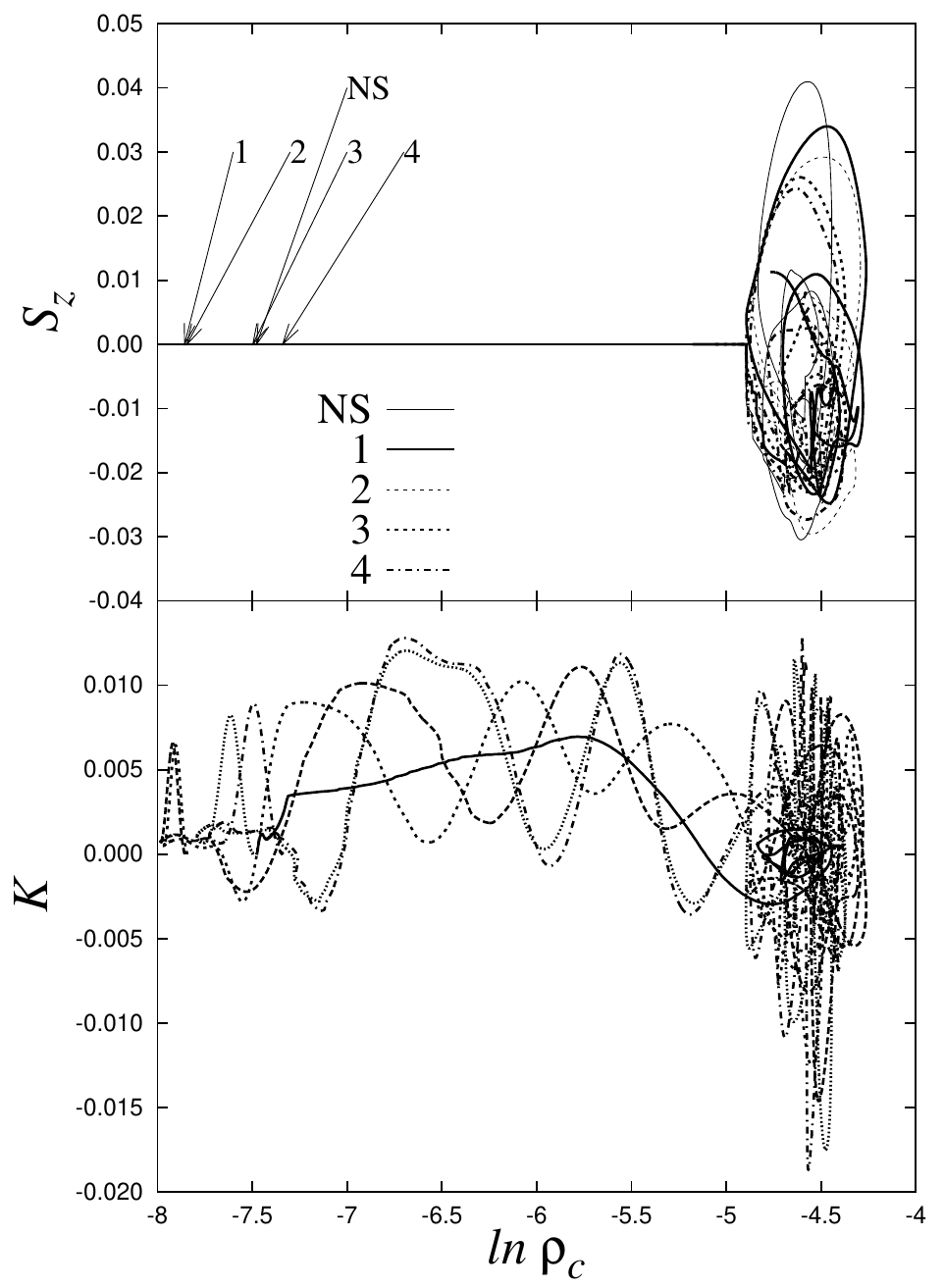}
\caption{The trajectories of all the configurations start on the left part and are attracted to the attracting basin on the right
part. The arrows point to the differing starting points of the trajectories. $K$ denotes the trace of the
extrinsic curvature at the coordinate (2.1,0.06,0.06) and $S_z$ denotes the z-component of the 3-momentum, $S_i$, of the fluid element
on $7.5\times 10^{-3}$ density contour, respectively. The natural logarithm is taken of the density at the center of the grid, namely
$ln \rho_c$, so as to greater distinguish the differing starting points of the trajectories in the phase diagrams.}
\end{figure}
Table 1 and Fig. 4 show the configurations that are constructed and their density distributions along the direction of collision
respectively. In Fig. 5, we draw the phase trajectories for these configurations using the density at the center of the grid, the trace 
of the extrinsic curvature at the coordinate (2.1,0.06,0.06) and the z-component of the 3-momentum, $S_i$, of the fluid element on the 
edge of the density contour bounded from below by $7.5\times 10^{-3}$ and compare them with that for the NS configuration in \cite{Jin} 
(the next section contains the reasons why these evolution variables are chosen for the phase diagrams).
We observe that these configurations are all attracted to the same attraction basin in the phase diagrams as that of the NS critical
solution. When the Gaussian heights exceed the critical values for each of the Gaussian packet configurations, NS-like objects are 
formed, while heights lower than the critical values produce black holes. This indicates the universality of the NS critical solution 
with respect to a 1-parameter family of initial data, and that the NS attraction basin constitutes a universal attractor.
\begin{figure}
\includegraphics[scale=0.85]{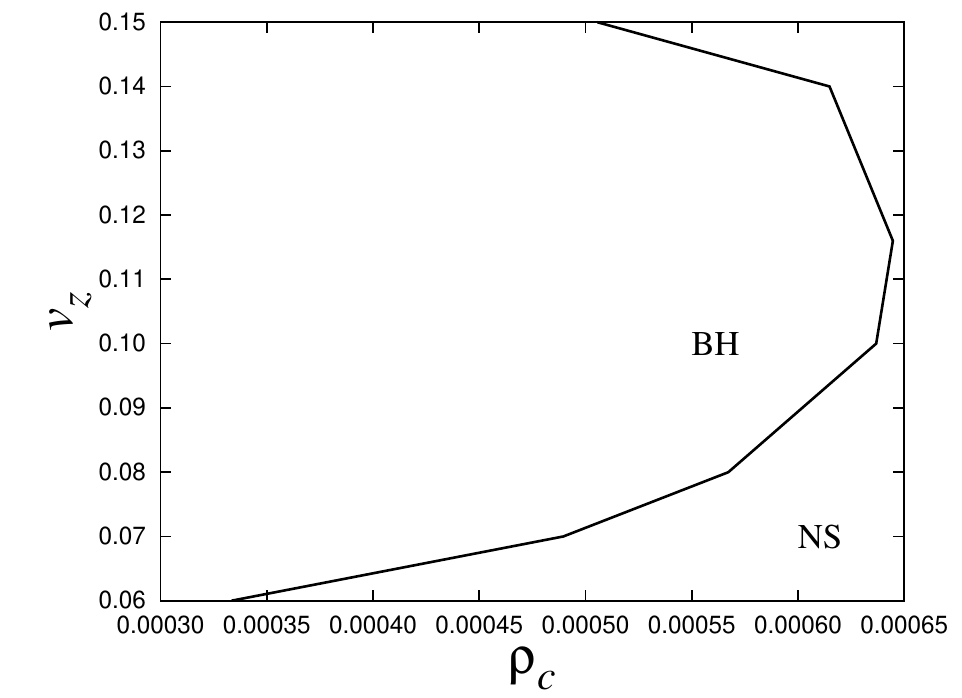}
\label{fig:6}
\caption{}
\end{figure}
\begin{figure}
  \begin{center}
    \begin{tabular}{cc}
      \scalebox{0.8}{\includegraphics{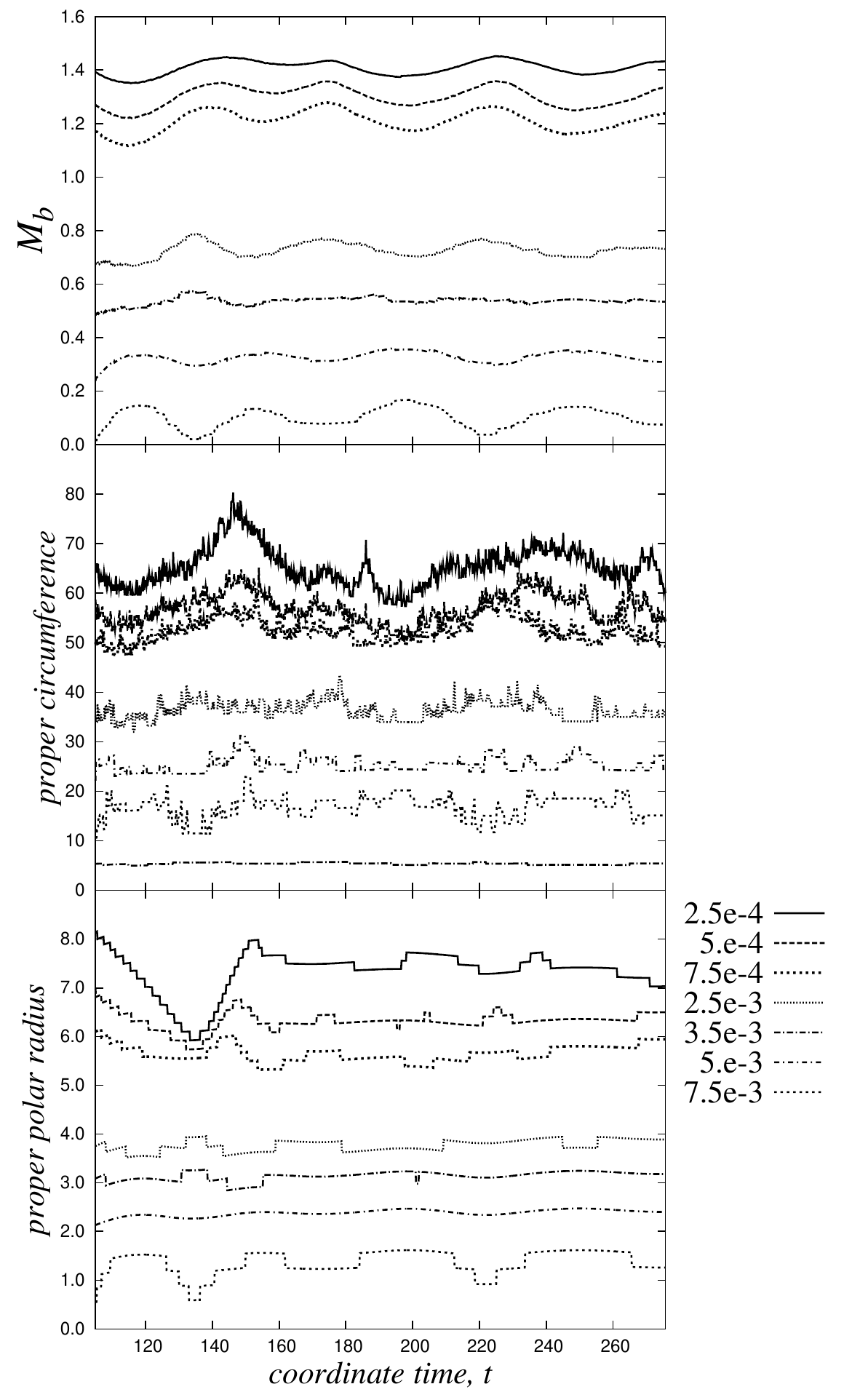}} 
      \scalebox{0.8}{\includegraphics{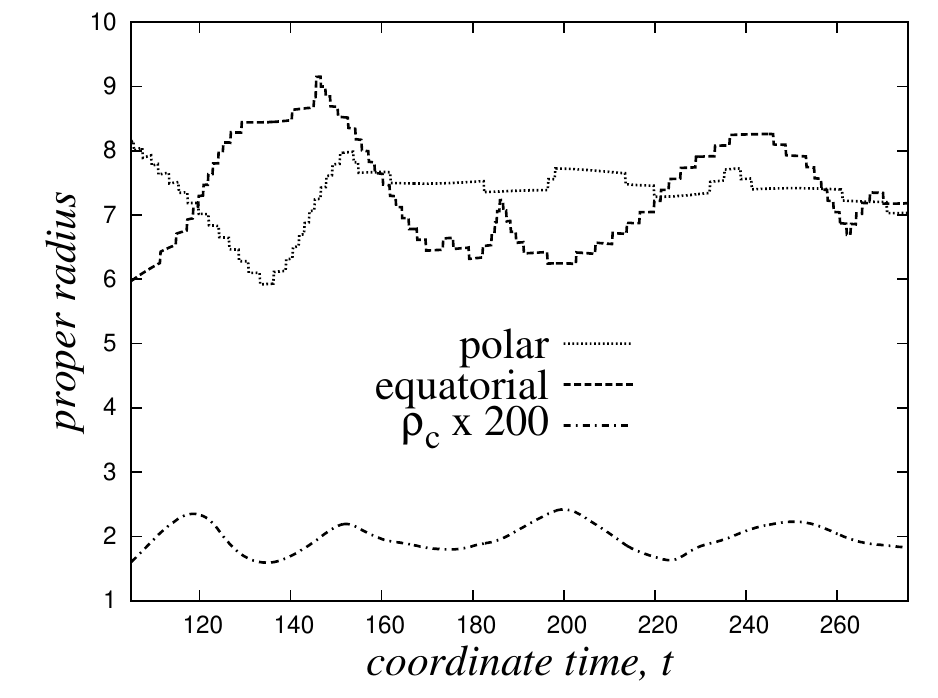}} 
    \end{tabular}
    \caption{Left legend shows contour densities at which measures are performed.
Polar and equatorial proper radii shown on the right are measured at the $5.0 \times 10^{-4}$ density contour. The polar 
direction is taken to be the z-axis on the xz-plane of the grid and the equatorial direction is taken to be the x-axis on the same 
plane. The equatorial direction is thus along the direction of collision of the NSs.}
    \label{7}   
  \end{center}
\end{figure}

\paragraph*{\bf Sec.4. Phase Space Analysis.}

In this section, we present a probe on the structure of the critical solution in a phase space framework. We note that the initial 
data configurations as presented in the previous section provide us with another degree of freedom in probing the structure of the 
critical collapse threshold in a more global scale. For this purpose, we vary the boost velocity of the Gaussian packets together with 
the Gaussian heights and widths while maintaining the baryonic mass of the system at $M_b=1.5393$. In Fig. 6, we plot boost velocity 
with respect to Gaussian height. We observe that there is a turning point in this phase diagram at approximately 
$(\rho_c,v)=(0.00064,0.12)$, which corresponds to the NS critical solution. An initial data configuration to the left of the 
critical surface collapses to a black hole and one to the right results in an NS-like object. This critical surface is an 
indication of the NS attraction basin as commented on in Sec. 3. Beyond both edges of the critical surface, 
no critical collapse behavior is observed. We therefore note that configurations with Gaussian packets with compactness outside the 
range of $0.00033\lesssim \rho_c\lesssim 0.00064$ are not attracted to the critical surface at all. However, due to the parabolic structure 
of this critical surface, configurations that are less compact than that with $\rho_c\sim 0.00064$ are more prone to undergo critical 
collapse than those that are more compact. We therefore note that not only the baryonic mass, but also the mass distribution, contributes to 
the dynamical mechanism that drives critical collapse. 

This is further reinforced by the observation of the presence of a stationary envelope surrounding the core of the merged object 
which undergoes critical collapse. For the NS case, we measure the baryonic mass enclosed within different density contours throughout 
the merged object, the proper polar circumferences of these density contours, and their equatorial and proper radii, and observe the 
pattern of their evolution throughout the critical collapse. The measure of the baryonic mass enclosed within different density contours 
throughout the merged object has the advantage of being a fully-geometric measure. In Fig. 7, we observe that this measure exhibit 
oscillations that are correlated between different regions of the merged object, whereas the corresponding proper polar circumferences 
remain basically stationary throughout the critical collapse phase ie. matter fluxes exist through the density contour with density threshold 
$\rho \sim 3.5 \times 10^{-3}$. The proper radii however exhibit prolonged oscillations only within the region bounded by this 
specific density contour. The oscillations of these radii outside this region exhibit damping into a stationary state. These 
oscillations are correlated between the polar and equatorial directions ie. the polar proper radius is in 
phase with the oscillation of the central density while the equatorial radius is out of phase. 

We also note that the oscillation characteristic is evident across the NS and Gaussian packet critical collapses. Further, the oscillation 
phases of certain evolution variables (as those mentioned in Sec. 3) correlate with each other such that they form closed periodic 
orbits in the phase diagrams, similar to the circle in the phase diagram of a simple pendulum. In infinite-dimensional 
dynamical systems theory \cite{Guckenheimer}, we note that phase space trajectory flows are tangent to invariant sets which can be 
divided into 3 types, ie. the center, stable and unstable subspaces. Stable manifolds contain flows that move toward the center 
manifold whilst unstable manifolds contain flows that move away. These manifolds are tangent to their respective subspaces. As a direct 
analogy, the critical collapse threshold corresponds to a $\Re^{(n-1)}$ center manifold that is bounded by stable and unstable $\Re^n$ 
manifolds. When a parameter of the initial data is varied, bifurcation occurs and solutions are carried away from the center manifold 
via the unstable manifold. Since stable manifolds definitionally contain flows that move toward the center manifold, bifurcation 
definitionally could not occur across stable manifolds. In both the NS and Gaussian packet cases, bifurcation is seen to occur across 
an oscillatory threshold. This strongly suggests that the center manifold in these cases contain a limit cycle rather than a fixed 
point.

\paragraph*{\bf Sec.5. Conclusions.}

In this work, we confirm that NS-like objects undergoing critical collapse are in a non-equilibrium state and therefore, their 
oscillations cannot be described by a perturbative mode analysis on equilibrium background space-times. The oscillation frequencies of 
these objects are found to be of 1 and 2 orders of magnitude smaller than that for equilibrium configurations. We note that the time scales 
of these oscillations, $T_{1}=0.21ms$ and $T_{2}=0.13ms$, as well as the time scale of solutions approaching and 
leaving the critical collapse threshold, $t=0.05ms$, are also of 1 order of magnitude smaller than the cooling time scales of NSs. This 
suggests a high possibility that NS systems undergoing a slow softening of the EOS, accretion or angular momentum loss undergo critical 
collapse. Critical phenomena in gravitational collapses of real astrophysical systems of NSs will thus have an effect on the emission of 
gravitational radiation that can be detected by LIGO, LISA etc. Work to investigate critical collapses in rotating NS-like objects, which 
greater resemble real NS systems, will be reported elsewhere by the authors KJJ and WMS.

We also confirm, in this work, that the NS critical solution constitutes a universal attraction basin for 1-parameter families of 
NS-like initial data. The unstable mode carrying solutions away from this critical collapse threshold is characterized by the complex   
pair of eigenvalues $\pm 0.09$. We present the critical surface resulting from bifurcations of a 1-parameter family of NS-like initial 
data and deduce that less compact mass distributions of NS-like objects are more prone to undergo critical collapses than more compact ones. 
With phase diagrams constructed from geometrically-invariant quantities and evolution variables of the 3+1 BSSN formulation using 
$\Gamma$-freezing shift and 1+log slicing, we present strong evidence that the NS critical collapse attractor is a limit cycle rather 
than a fixed point. Since an exact critical solution is by definition non-radiative, the presence of a limit cycle on the NS 
critical collapse threshold together with the observation of critical collapse in rotating NS-like systems, poses an interesting 
question of how a gravitational system maintains oscillations and rotations without emitting gravitational radiation. 
Further investigations of this question will be closely related to perturbative mode analyses on non-equilibrium configurations.

We thank K. Young for helpful discussions. Numerical computations are performed on WUGRAV machines and NCSA clusters. 
The research is supported by the McDonnell Center for the Space Sciences, Washington University, and the NSF Grant MCA93S025.
\bibliographystyle{prsty} 
\bibliography{references}

\end{document}